\documentclass[8pt,a4paper,twocolumn]{article}
\usepackage[margin=20mm]{geometry}
\usepackage{cite}
\usepackage{amsmath,amssymb,amsfonts}
\usepackage[ruled]{algorithm2e} 
\usepackage{graphicx}
\usepackage{textcomp}
\usepackage{xcolor}
\usepackage{url}
\usepackage{hyperref}
\usepackage{tabularx}
\usepackage{colortbl,array}
\usepackage{booktabs}
\hypersetup{pdfborder={0 0 0}}
\newcolumntype{C}{>{\columncolor{blue!10}}c}
\newcolumntype{R}{>{\columncolor{blue!10}}r}
\def\BibTeX{{\rm B\kern-.05em{\sc i\kern-.025em b}\kern-.08em
　　T\kern-.1667em\lower.7ex\hbox{E}\kern-.125emX}}
\begin{document}

\title{\Huge Correlation-diversified portfolio construction by finding maximum independent set in large-scale market graph}

\author{
Ryo Hidaka$^{\ast}$, Yohei Hamakawa, Jun Nakayama, \\
and Kosuke Tatsumura \\
\small Corporate Research and Development Center, Toshiba Corporation, Japan\\
$^{\ast}$\small Corresponding author: Ryo Hidaka (e-mail: ryo.hidaka@toshiba.co.jp)
}
\date{}

\maketitle

\begin{abstract}
Correlation-diversified portfolios can be constructed by finding the maximum independent sets (MISs) in market graphs with edges corresponding to correlations between two stocks. The computational complexity to find the MIS increases exponentially as the size of the market graph increases, making the MIS selection in a large-scale market graph difficult.
Here we construct a diversified portfolio by solving the MIS problem for a large-scale market graph with a combinatorial optimization solver (an Ising machine) based on a quantum-inspired algorithm called simulated bifurcation (SB) and investigate the investment performance of the constructed portfolio using long-term historical market data.
Comparisons using stock universes of various sizes [TOPIX~100, Nikkei~225, TOPIX~1000, and TOPIX (including approximately 2,000 constituents)] show that the SB-based solver outperforms conventional MIS solvers in terms of computation-time and solution-accuracy. By using the SB-based solver, we optimized the parameters of a MIS portfolio strategy through iteration of the backcast simulation that calculates the performance of the MIS portfolio strategy based on a large-scale universe covering more than 1,700 Japanese stocks for a long period of 10 years. 
It has been found that the best MIS portfolio strategy (Sharpe ratio = 1.16, annualized return/risk = 16.3\%/14.0\%) outperforms the major indices such as TOPIX (0.66, 10.0\%/15.2\%) and MSCI Japan Minimum Volatility Index (0.64, 7.7\%/12.1\%) for the period from 2013 to 2023.
\end{abstract}

\section{Introduction}\label{sec:introduction}
From the establishment of Modern Portfolio Theory (MPT) by Harry Markowitz and William F. Sharpe in the mid-20th century to the present~\cite{markowitz1952, sharpe1964}, improving the risk-return characteristics of portfolios by diversifying across multiple financial instruments has remained an important issue for both institutional and individual investors.
If portfolios can be constructed with stocks that have low correlations to each other, the price movements of the component stocks will be disparate and their contributions to the overall portfolio performance will cancel each other out, thus reducing the risk of the portfolio.
Such a correlation effect is one of the basis of the MPT.
In the last decade or two, attention has focused on risk-based portfolios and minimum volatility indices~\cite{koumou20}, and a variety of indices and funds linked to them have been launched.
There are also a lot of studies on the portfolios that focus on risk, such as ones using  genetic algorithm~\cite{chou2017} and deep reinforcement learning~\cite{jin2023}.

The ``market graph,'' which (graphically) represent the correlation of stocks, has been actively studied~\cite{butenko03, boginski04, boginski2005statistical, kalyagin2014market, marzec2016portfolio}.
The market graph is an undirected graph with stocks as nodes, connected by edges between nodes (stocks) with a certain level of correlation; in other words, its graph structure (connection relationship) represents the correlation of the entire stock universe.
A portfolio constructed with unconnected nodes on the market graph, or an independent set, will be a diversified portfolio with low correlations to each other, which is expected to reduce risk.
Kalyagin \textit{et al.}~\cite{kalyagin2014market} investigated the relationship of independent sets of market graphs to Markowitz's portfolio theory and demonstrated the usefulness of the market graph.

The independent set of a market graph with the maximum number of nodes is obtained by solving the maximum independent set (MIS) problem, one of the combinatorial optimization problems.
While the MIS problem has a variety of practical applications~\cite{butenko2002finding, lee2006post, zhou2017efficient, mallick2020using}, it is known to be nondeterministic polynomial time (NP)-hard, hence the computational complexity increases exponentially as the number of nodes (stocks in the market graph) increases.
Butenko~\cite{butenko03} used a greedy heuristic to solve the MIS problem for market graphs in a reasonable time.

Recently, Ising machines~\cite{sbm1, FPL19, sbm2, NatEle, kanao2022simulated, johnson11, king23, honjo21, pierangeli19, cai20, aadit22, moy22, sharma22, takemoto19, kawamura23, yamamoto21, matsubara20, waidyasooriya21, okuyama19}, which specialize in obtaining high-accuracy solutions of combinatorial optimization problems in short times, have attracted much attention.
The Ising machine is a device that searches for the ground-state of the Ising model that is a mathematical model of magnetic materials (this search is called the Ising problem), and many combinatorial optimization problems, including MIS problems, can be mapped onto Ising problems~\cite{lucas14}.
Yarkoni \textit{et al.}~\cite{yarkoni18} solved a random MIS problem with D-Wave's quantum annealing (QA) processor~\cite{johnson11, king23}, an Ising machine based on quantum principles, and reported speed and solution-accuracy advantages over conventional MIS solvers at graph sizes of up to 40 nodes.
However, the number of nodes (stocks) of 40 is not sufficient for practical portfolio selection, as it requires handling thousands of stocks when considering the whole constituents of a market index as the universe for stock selection [e.g. the Tokyo Stock Price Index (TOPIX) as the proxy for the Japanese stock market involves more than 2,000 stocks, and the NASDAQ Composite Index for the NASDAQ market involves more than 3,000 stocks].

Simulated bifurcation (SB)~\cite{sbm1} is a quantum-inpsired heuristic algorithm for combinatorial optimization. The algorithm of SB is highly parallelizable and enables developing the massive-parallel accelerators (Ising machines) at a large scale (including more than several thousand Ising spins) with parallel processors such as field programmable gate arrays (FPGAs)~\cite{sbm1,FPL19, sbm2, NatEle} and graphics processing units (GPUs)~\cite{sbm1, sbm2}. SB was derived from a classical counterpart to a quantum adiabatic optimization method called quantum bifurcation~\cite{qbm}, and variations of SB have been proposed: adiabatic SB (aSB) ~\cite{sbm1}, ballistic SB (bSB) ~\cite{sbm2}, discrete SB (dSB) ~\cite{sbm2}, and heated SB~\cite{kanao2022simulated}. The hardware implementations of SBs are called simulated bifurcation machines (SBMs). Financial automated trading systems using SBMs for detecting short-lived trading opportunities have been reported~\cite{tatsumura20, tatsumura23a, tatsumura23b}.

In this work, we build an FPGA-based accelerator for bSB (hereafter, the SBM), and by using it, investigate the performance of a MIS portfolio strategy based on the market graphs covering over 1,700 Japanese stocks when varying the strategy parameters with iterating the 10-year backcast simulation. The performance of the best MIS portfolio strategy is compared with the Japanese major indices from the perspective of risk-return characteristics.

The SBM supports fully-connected 2,048-spin Ising models and is capable of solving 2,048-node MIS problems, which is implemented as a peripheral component interconnect-express (PCIe)-attachable, look-aside type, FPGA-based acceleration card to the CPU-based host system.
The backcast simulation is carried out with the CPU, but the computationally-hard parts, namely the parts to solve the MIS problems, are offloaded to the SBM without large network-communication overheads for Web APIs.
We also developed the hardware abstraction layer (HAL) for the SBM and APIs (Application Programming Interface) for C/C++ and Python programming languages, making the SBM accessible to the backcast simulator.

To evaluate SBM performance as a MIS solver, we compare it with conventional MIS solvers, NetworkX~\cite{hagberg2008exploring} and OR-tools~\cite{ortools}, in terms of computation-time and solution-accuracy when solving the MIS problems of market graphs representing the constituents of Japanese stock indices (TOPIX, TOPIX~1000, Nikkei~225, and TOPIX~100).

We repeat the backcast simulations of the MIS portfolio strategy with TOPIX as the stock universe (over 1,700 constituent stocks) for the period from 2013 to 2023 under a total of 38 parameter settings (a combination of 19 patterns of thresholds that determine the edge connections of the market graphs and two different asset-allocation methods for the selected portfolios).
And then, we compare the best MIS portfolio strategy with the major indices such as TOPIX and MSCI Japan Minimum Volatility Index, and furthermore analyze its risk-return characteristics.

The rest of the paper is organized as follows.
Section~\ref{sec:Portfolio strategy} describes the MIS portfolio strategy and the formulation.
Section~\ref{sec:System architecture} details the MIS portfolio simulator and the interface and core circuit of the SBM. 
Section~\ref{sec:MIS solver performance} compares the SB-based MIS solver with the conventional MIS solvers.
Section~\ref{sec:Portfolio performance} systematically shows the performance of  the portfolio strategy when varying the strategy parameters and analyzes the property of the best MIS strategy.
Section~\ref{sec:Conclusion} concludes the paper.

\begin{figure}[b]
\centering
\includegraphics[width=8.3 cm]{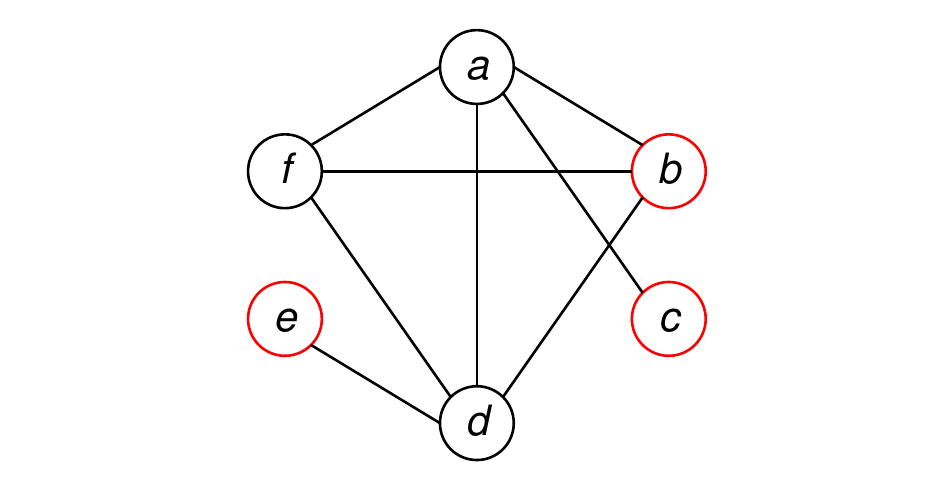}
\caption{Market graph for an $N$-stock universe ($N = 6$). Red nodes are the maximum independent set (MIS).}
\label{Fig_MIS_problem}
\end{figure}

\section{Portfolio strategy}\label{sec:Portfolio strategy}
\subsection{Portfolio construction \& management}\label{sec:Portfolio construction and management}
\subsubsection{Portfolio selection}\label{sec:Portfolio selection}
The proposed strategy selects a portfolio of stocks from a stock universe (a group of tradable stocks) according to the maximum independent set (MIS) found in the market graph representing the stock universe. The market graph is an undirected graph with nodes corresponding to the stocks in the universe and each edge corresponding to the presence of the correlation between the two nodes (/stocks) connected by the edge. The definition of the market graph and the constituents of the stock universes examined in this work will be detailed in Sec.~\ref{sec:Definition of market graph} and Sec.~\ref{sec:Stock universe}, respectively. Fig.~\ref{Fig_MIS_problem} shows an example of the market graph for an $N$-stock universe ($N = 6$) and the MIS in there. The solution to the MIS problem is a binary vector, where $i$th binary value means whether the $i$th stock is involved in the MIS. In this strategy, we choose the stocks involved in the MIS as the constituents of the portfolio.

\subsubsection{Portfolio composition}\label{sec:Portfolio composition}
After the portfolio selection, we determine the portfolio composition, i.e., the asset allocation to each constituent of the portfolio. The weight $w_{i}$ represents the ratio of the asset allocated for the stock $i$ to the total asset. 
We examine the following two asset-allocation methods.

A) Equal Weight (EW): All constituents have the same weights; the weight $w_{i}$ of stock $i$ is represented by
\begin{equation}\label{Eq_EW}
w_{i}=\frac{1}{N}.
\end{equation}

B) Inverse Volatility Weight (IVW): The weight is proportional to the inverse of the volatility of each constituent; $w_{i}$ is defined by
\begin{equation}\label{Eq_IVW}
w_{i}=\frac{v_{i}^{-1}}{\sum\limits_{k=1}^{N} v_{k}^{-1}},
\end{equation}
where the volatility $v_{i}$ is the standard deviation of the logarithmic daily return [$R_{i}(t)$]~\cite{miskolczi2017note}. The $R_{i}(t)$ is expressed by
\begin{equation}\label{Eq_Ri}
R_{i}(t)=\ln \frac{P_{i}(t)}{P_{i}(t-1)},
\end{equation}
where $P_{i}(t)$ is the closing price of stock $i$ on day $t$ (the last traded price on the business day) and the dividend (if exists) is considered. The standard deviation of $R_{i}(t)$ over the last three years is the volatility $v_{i}$.

\subsubsection{Portfolio rebalance}\label{sec:Portfolio rebalance}
The portfolio is rebalanced monthly; the portfolio selection and weighting are carried out every month (at the end of each month). For the difference between the new portfolio and the last-month portfolio in terms of the constituents and composition, we take into account the trading costs corresponding to 0.1\% of the total amount of transactions (the sum of the amount of shares purchased and the absolute amount of shares sold).
The evaluation value of a portfolio in month $t$, $P_\mathrm{PF}(t)$, is the sum of the evaluation values of constituents in the portfolio, where the evaluation value of a constituent is the product of the base price and share at the end of the month. The monthly return of the portfolio strategy, $R_\mathrm{PF}(t)$, is expressed by
\begin{equation}\label{Eq_RPF}
R_\mathrm{PF}(t)=\frac{P_\mathrm{PF}(t)}{P_\mathrm{PF}(t-1)}-1.
\end{equation}

\subsection{Finding MIS in market graph}\label{sec:Finding MIS in market graph}
\subsubsection{Definition of market graph}\label{sec:Definition of market graph}
We follow the definition and generation procedure of the market graph in Ref.~\cite{butenko03}.
In the correlation matrix $C$, the correlation coefficient $C_{i,j}$ between stocks $i$ and $j$ is calculated for the last $T$ days as follows.
\begin{equation}\label{Eq_Cij}
C_{i,j}=\frac{\sum\limits_{t}^{T} (R_{i}(t)-\overline{R_{i}})(R_{j}(t)-\overline{R_{j}})} {\sqrt{\sum\limits_{t}^{T} (R_{i}(t)-\overline{R_{i}})^2}\sqrt{\sum\limits_{t}^{T} (R_{j}(t)-\overline{R_{j}})^2}}.
\end{equation}
$R_{i}(t)$ is the logarithmic daily return defined by Eq.~\ref{Eq_Ri}. The average of $R_{i}(t)$ for the last $T$ days, $\overline{R_{i}}$, is given by 
\begin{equation}\label{Eq_Ri_bar}
\overline{R_{i}}=\frac{1}{T} \sum\limits_{t}^{T} {R_{i}(t)}.
\end{equation}
Upon the portfolio selection at the beginning of each month, we recalculate the correlation matrix $C$ using historical price data for the last three years (the $T$ days are the business days involved in the last three years). The correlation coefficient $C_{i,j}$ varies from -1 to 1.
We can also consider $C_{i,j}$ as the similarity of the price fluctuations of the stocks $i$ and $j$, i.e., the greater the value, the more similar the price movements of the two stocks.

The connectivity (or topology) of the market graph is determined by the correlation matrix $C$ and a threshold parameter of $\theta$; the nodes $i$ and $j$ in the market graph are connected if the correlation coefficient $C_{i,j}$ is equal to or greater than $\theta$. The connectivity of the market graph is sensitive to $\theta$ and the choice of $\theta$ characterizes the performance of a MIS portfolio strategy. In this work, we determine the best $\theta$ by analyzing the long-term performance of the MIS portfolio strategy for one of various $\theta$ and then assume using a fixed $\theta$ for the MIS strategy (not changed monthly).

\subsubsection{QUBO formulation of MIS problem}\label{sec:QUBO formulation of MIS problem}
We describe the MIS problem in the form of quadratic unconstrained binary optimization (QUBO) according to Ref.~\cite{lucas14}. The QUBO problem is mathematically equivalent to the Ising problem. Considering a graph $G = (V, E)$ consisting of a node set $V$ and an edge set $E$, the connection $f_{i,j}$ between nodes $i$ and $j$ is represented as follows.
\begin{equation}\label{Eq_f}
f_{i,j}=
\left\{ 
\begin{alignedat}{2} 
1 & \;\;(i,j \in E), \\
0 & \;\;(i,j \not\in E).
\end{alignedat} 
\right.
\end{equation}

In the QUBO formulation, we search for the bit configuration $\{b_{i}\}$ that minimizes the cost function $H$. The cost function $H$ is designed to decrease with the number of nodes selected (for maximizing the independent set) and increase if connected nodes with edges are selected (if the connected nodes are selected, the subset selected is no longer an independent).
The $\{b_{i}\}$ and $H$ are defined as follows.
\begin{equation}\label{Eq_bi}
b_{i}=
\left\{ 
\begin{alignedat}{2} 
1 & \;\;(\text{Node $i$ is an element of the independent set}), \\
0 & \;\;(\text{Node $i$ is not an element of the independent set}),
\end{alignedat} 
\right.
\end{equation}
\begin{align}\label{Eq_H}
H&=H_{A}+H_{B},\\[10pt]
H_{A}&=A\sum_{i,j}^{} f_{i,j}b_{i}b_{j},\\[10pt]
H_{B}&=-B\sum_{i}^{} b_{i}.
\end{align}
$H_{A}$ increases (gives a penalty) when a connected node is selected and is minimized when the subsets are formed only with unconnected nodes.
$H_{B}$ is proportional to the number of nodes selected; the higher the number, the lower the cost.
$A$ and $B$ are positive coefficients and must be $B < A$ to avoid decreasing the cost function despite constraint violations (selecting connected nodes). In this work, A = 2 and B = 1.

\begin{figure}[t]
\centering
\includegraphics[width=8.3 cm]{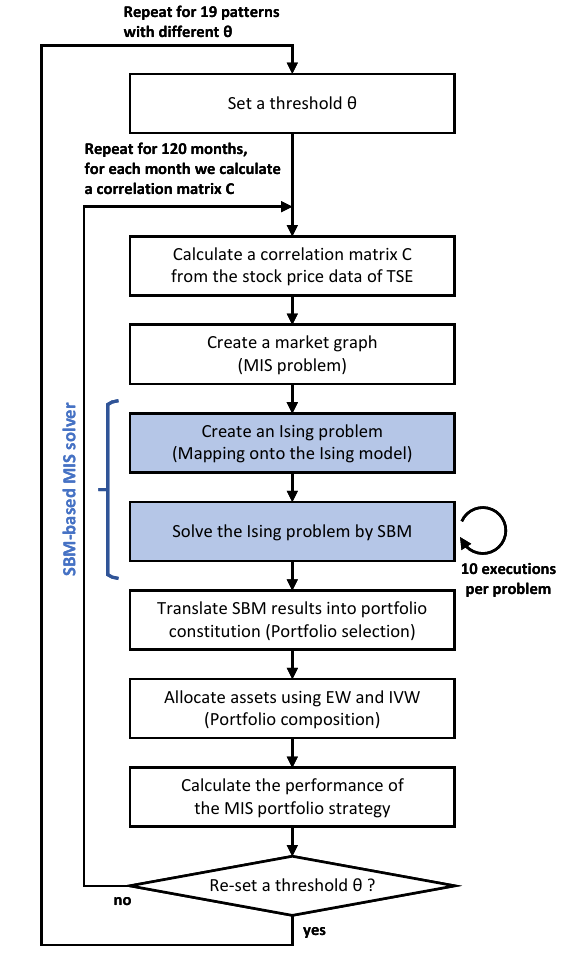}
\caption{Flowchart for simulation of MIS portfolio strategy.}
\label{Fig_MIS_flowchart}
\end{figure}

\subsubsection{Stock universe}\label{sec:Stock universe}
In the following sections (Sec.~\ref{sec:MIS solver performance} and Sec.~\ref{sec:Portfolio performance}), we use stock universes of various sizes corresponding to several market indices (TOPIX~100, Nikkei~225, TOPIX~1000, and TOPIX) for the Tokyo Stock Exchange (TSE). See APPENDIX~\ref{sec:Market indices in the TSE} for the details on the market indices.

Since the constituents of market indices sometimes change, we determined the constituents of the stock universes as follows. The candidates of the stock universe for each market index are the constituents of the market index as of May 2023. From the candidates, the stocks that have remained listed for the analysis period consist of the stock universe. The stock universes corresponding to TOPIX in Sec.~\ref{sec:MIS solver performance} is 2,026, but that in Sec.~\ref{sec:Portfolio performance} is 1,747. This difference comes from the difference in the analysis periods; the analysis period in Sec.~\ref{sec:MIS solver performance} is 2017 to 2023, while that in Sec.~\ref{sec:Portfolio performance} is 2010 to 2023.

\begin{figure}[t]
\centering
\includegraphics[width=8.3 cm]{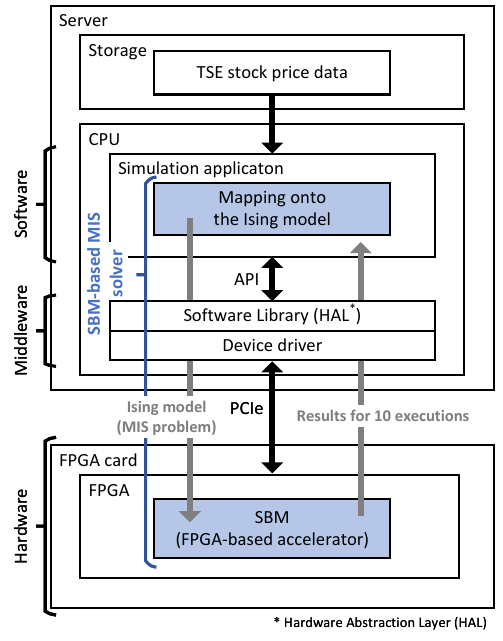}
\caption{System architecture of the MIS portfolio simulator.}
\label{Fig_MIS_system}
\end{figure}

\subsection{Portfolio simulation}\label{sec:Portfolio simulation}

Figure~\ref{Fig_MIS_flowchart} shows the flowchart for the simulation of the MIS portfolio strategy. In Sec.~\ref{sec:Portfolio performance}, in order to determine the best $\theta$ and asset-allocation method, we repeat the 10-year (120 months) backcast simulation while changing $\theta$ (19 patterns) for the constituents of TOPIX as the universe and then calculate the annualized return and risk from the 10-year simulation data. 

At the beginning of the simulation, $\theta$ is set and then the simulation evaluates the performance of the strategy every month. At the beginning of the monthly processing, the correlation matrix $C$ and the volatility $\{v_{i}\}$ are calculated based on the price data for the last three years before the month as described in Sec.~\ref{sec:Definition of market graph} and Sec.~\ref{sec:Portfolio composition}. The market graph is constructed from the correlation matrix $C$ by determining edge connections with the $\theta$ and then the Ising problem to be solved is created from the market graph (mapping of the MIS problem for the market graph onto the Ising problem) as described in Sec.~\ref{sec:QUBO formulation of MIS problem}. The SBM solves the Ising problem 10 times with different initial states and obtained 10 different solutions. Note that SB is a heuristic algorithm and can output different solutions by changing the initial states. After verifying the solutions in terms of constraint violation (if connected nodes are included), we select the best solution having the largest size of the independent set from the verification-passed solutions. The best solution is translated directly into the portfolio constitution (portfolio selection) and then the compositions of the portfolio are determined for the two cases of the asset allocation methods (EW and IVW). Finally, we evaluate the new portfolio (with considering the trading costs) and calculate the monthly return by comparing the new portfolio with the last-month portfolio as described in Sec.~\ref{sec:Portfolio rebalance}.

\section{System architecture}\label{sec:System architecture}
\subsection{Portfolio simulator}\label{sec:Portfolio simulator}

Figure~\ref{Fig_MIS_system} shows the architecture of the MIS portfolio simulator which executes the simulation process in the flowchart (Fig.~\ref{Fig_MIS_flowchart}) using the historical stock price data of the TSE. The simulator physically consists of a CPU-based server and an FPGA card installed in the PCIe slot of the server, which is logically modeled as three layers of software, middleware, and hardware. See APPENDIX~\ref{sec:Implementation details} for the details of the hardware used.

The SBM (hardware layer) is implemented as a massively-parallel custom circuit with the FPGA and the server as a look-aside accelerator (offloader) to the CPU-based host system. In the flowchart, the computationally-hard part, namely solving the NP-hard Ising problem (MIS problem), is offloaded to the SBM (dedicated accelerator) to reduce the overall simulation time. The remaining in the flowchart is executed by CPU processing (software layer). To solve the MIS problem in the monthly market graph by the SBM (an Ising machine), we have to map the MIS problem onto the Ising model. The mapping is executed also by CPU. Thus, the SBM-based MIS solver (blue highlighted in Figs.~\ref{Fig_MIS_flowchart} and~\ref{Fig_MIS_system}) consists of mapping the MIS problem and solving the Ising problem, unlike the other software MIS solvers that directly solves the MIS problems. In the performance comparison of the MIS solvers (Sec.~\ref{sec:MIS solver performance}), we include the times of mapping and communication needed for the SBM-based solver in the computation-time.

To make the SBM accessible at a higher abstraction level to the simulator program running on the CPU, we developed the middleware layer, namely, the hardware abstraction layer (HAL) for the SBM and the APIs (Application Programming Interface) for C/C++ and Python programming languages. Seen from the CPU side, the HAL abstracts the detailed functionality of the device driver for the SBM. The APIs receive the SBM execution parameters such as the input data describing the Ising problem and the number of times to consecutively executes the SBM (10 times in this work), and then return the results (solutions) for the executions to the simulation program running in the software layer. More precisely, the API function calls the functions of the device driver to control PCIe communication between the CPU and the FPGA card and run the SBM core circuit. The SBM solves the Ising problem based on the execution parameters received via the PCIe communication with different initial states generated by the internal random number generator and then transmits the results (for 10 executions) back to the host system.

\subsection{SBM core circuit}\label{sec:SBM core circuit}
We describe the core circuit architecture for the ballistic SB algorithm (bSB)~\cite{sbm2}. The overall circuit architecture for bSB follows the basic SBM design for adiabatic SB (aSB)~\cite{FPL19}, but the data path (pipeline path) for the time-evolution part is totally modified to represent the processing of the time-evolution producer specific to the bSB. In physical representation of the bSB, a perfectly inelastic wall has been introduced to the mechanical system of the bSB. The circuit architecture for bSB has not been illustrated in the paper that introduced the bSB algorithm~\cite{sbm2}.

\begin{figure}[t]
\centering
\includegraphics[width=8.3 cm]{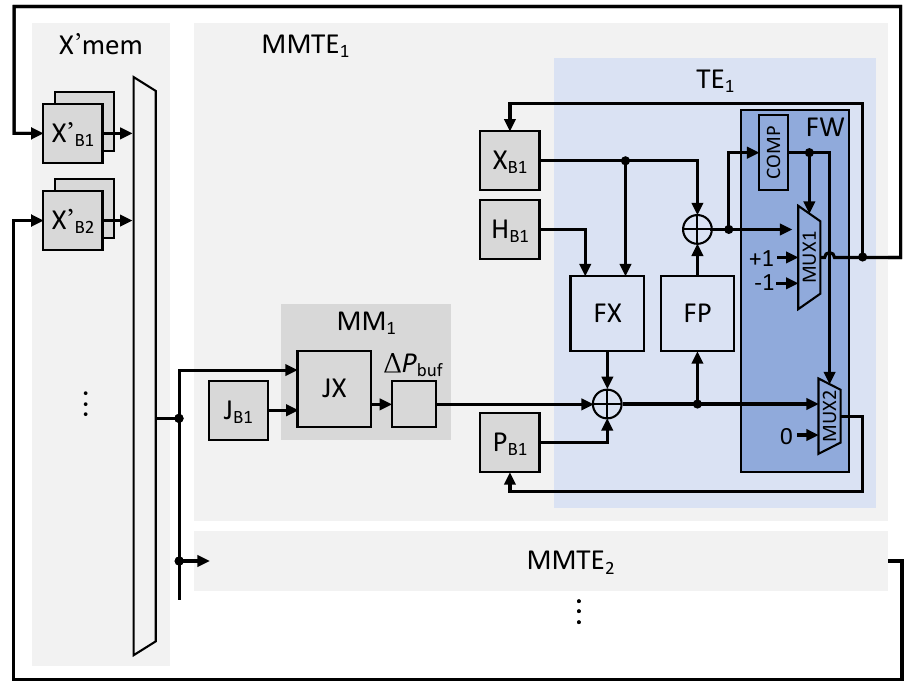}
\caption{Core circuit architecture of the bSB accelerator.}
\label{Fig_SBM_circuit}
\end{figure}

Simulated bifurcation~\cite{sbm1, sbm2} numerically simulates the time-evolution of $N$ nonlinear oscillators according to the Hamiltonian equations of motion, where the nonlinear oscillators correspond to the spin variables and the state of $i$th oscillator is described by the position and momentum ($x_{i}$, $p_{i}$). The SB time-evolution step consists of calculating the correction of momenta $\{\Delta p_{i}\}$ based on the many-body interaction [computationally corresponding to the matrix-vector multiplication (MM) of the $J$ coupling matrix and $\{ x_{i}\}$ position vector] and calculating the updated (time-evolved: TE) state variables, $\{ x_{i}^{k+1}\}$ and $\{ p_{i}^{k+1}\}$, from the $\{\Delta p_{i}\}$, bias $\{h_{i}\}$, and the current state variables, $\{ x_{i}^{k}\}$ and $\{ p_{i}^{k}\}$. After executing the time-evolution steps for a predetermined number ($N_{step}$), the positions of oscillators are digitized to be spins [the sign of $x_{i}$, sgn($x_{i}$), gives the state of $i$th spin].

Figure~\ref{Fig_SBM_circuit} shows the block diagram of the SB core circuit, where the pipeline path for the time-evolution part specific to the bSB is blue highlighted. The main two computation components are $\mathrm{MM}$ units corresponding to the multiply-accumulate (MAC) operations of $\sum_{j=1}^{N}J_{ij}x_{j}$ and $\mathrm{TE}$ pipelines corresponding to the time-evolution operation, which are combined to be $\mathrm{MMTE}$ units (each responsible for updating a subgroup of nonlinear oscillators). The $\mathrm{MMTE}$ units are organized with the global $\mathrm{X^{\prime}_{mem}}$ memory unit to make a circulative structure as a whole corresponding to the iteration of the SB time-evolution steps. Memory modules (connected to TE) $\rm{X_{B}}$, $\rm{P_{B}}$, and $\rm{H_{B}}$ store $x_{i}$, $p_{i}$, and $h_{i}$, respectively.

The feature of bSB is the perfectly inelastic walls existing at the positions of +1 or -1. If $i$th oscillator collides with the wall, the position and momentum of the oscillator are replaced by predetermined values [$x_{i}\leftarrow sgn(x_{i})$, $p_{i}\leftarrow 0$].
The TE pipeline consists of the FX (calculates the value to update $p_{i}$ based on $x_{i}$), FP (calculates the value to update $x_{i}$ based on $p_{i}$), and FW (corresponds to the wall) modules, where the functionalities are expressed as follows.
\begin{align}\label{Eq_FXFPFW}
FX(x_{i},h_{i})&=\Delta t\:\{-(\alpha_{0}-\alpha)x_{i}-\eta h_{i}\},\\[10pt]
FP(p_{i})&=\Delta t\:p_{i},\\[10pt]
FW(x_{i})&=
\left\{ 
\begin{alignedat}{2}
sgn(x_{i}) & \;\;(\text{when $|x_{i}|>1$}), \\
x_{i} & \;\;(\text{when $|x_{i}|\le1$}),
\end{alignedat} 
\right.\\[5pt]
FW(x_{i}, p_{i})&=
\left\{ 
\begin{alignedat}{2} 
0 & \;\;(\text{when $|x_{i}|>1$}), \\
p_{i} & \;\;(\text{when $|x_{i}|\le1$}),
\end{alignedat} 
\right.
\end{align}
where $\Delta t$, $\alpha_{0}$, $\alpha$ and $\eta$ are the SB parameters, the same as in the previous work~\cite{FPL19}. The conditional-branch processing of the FW consists of detecting the collision and processing depending on the branch, which are implemented, respectively, with one 32-bit comparator (COMP) and two 32-bit multiplexers (MUX1 and MUX2) as shown in Fig.~\ref{Fig_SBM_circuit}. 

The SBM implemented is capable of solving fully-connected 2,048-spin Ising problem with a computational precision of 32-bit floating-point, which works at the system clock frequency of 239 MHz in a highly-parallel fashion; the MM units have totally 4,096 MAC components (basic processing elements), which compute 4,096 spin-spin interactions in a single clock and there are 64 TE pipelines, including 64 sets of COMP, MUX1, and MUX2. The circuit resources used for the perfectly inelastic wall are not significant with compared to the MAC components. The parameters for solving the MIS problem are $N_{step}$ = 1000, $\Delta t$ = 0.2 and $\eta$ = 0.2.

\section{MIS solver performance}\label{sec:MIS solver performance}
The SBM-based MIS solver is evaluated in terms of computation-time and solution-accuracy through comparison with conventional MIS solvers based on heuristic and exact algorithms (NetworkX~\cite{hagberg2008exploring} and OR-tools~\cite{ortools}).

NetworkX~\cite{hagberg2008exploring} is an open-source Python library for analyzing and visualizing graph structures, including a heuristic solver for the MIS problem. The size of an approximate solution for the MIS problem in a graph $G = (V, E)$ with nodes $V$ and edges $E$ is $O(|V|/(\log |V|)^{2})$ in the worst case~\cite{boppana1992approximating}. OR-tools~\cite{ortools} provides a solver for 0-1 integer programming problems applicable to the MIS problem, which is based on a branch-and-cut algorithm~\cite{padberg1991branch} known as an exact-solution method. Hereafter, NetworkX and OR-tools refer to the heuristic and exact-solution solvers for the MIS problem above mentioned. See APPENDIX~\ref{sec:Implementation details} for the details of the execution.

\begin{figure}[h]
\centering
\includegraphics[width=8.3 cm]{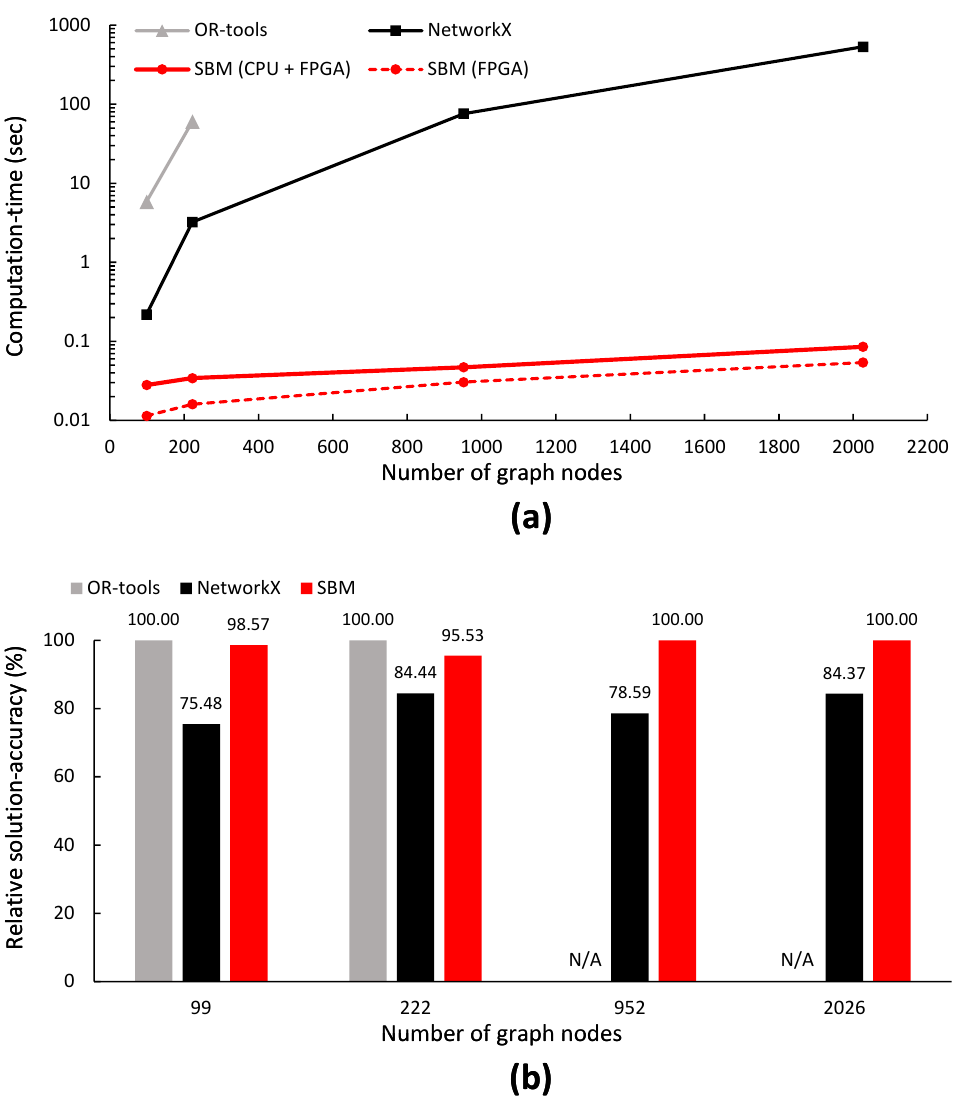}
\caption{Performance comparison of the three MIS solvers, SBM (heuristic), NetworkX (heuristic), and OR-tools (exact-solution), in terms of (a) computation-time and (b) relative solution-accuracy (the size of the independent set found). The computation-time (the shorter, the better) and the size of the independent set found (the larger, the better) when solving the MIS problem for each of 10 market graphs having the same size but different edges were measured and averaged over the 10 market graphs. Each value in (b) is the ratio to the largest one of the (averaged) sizes of independent sets found by the three solvers.}
\label{Fig_MIS_performance}
\end{figure}

Figure~\ref{Fig_MIS_performance} shows the comparison results of the SBM-based MIS solver (heuristic), NetworkX (heuristic), and OR-tools (exact-solution) in terms of computation-time and solution-accuracy. The four sizes of stock universes corresponding to the market indices (TOPIX~100, Nikkei~225, TOPIX~1000, and TOPIX) were used and 10 market graphs (having the same size but different edges) for each universe were generated with $\theta$ = 0.25 based on the correlation matrices for the different periods (consecutive three years, starting from different dates) in Mar. 2017 to May 2023. The numbers of nodes (corresponding to the number of the stocks in the universe) in the market graphs are 99, 222, 952, and 2,026 (see Sec.~\ref{sec:Stock universe}). The computation-time and the size of the independent set found (the larger, the better) when solving the MIS problem for each of the 10 market graphs were measured and averaged over the 10 market graphs. As mentioned in Sec.~\ref{sec:Portfolio simulator}, the computation-time per market graph for the SBM [the red solid line in Fig.~\ref{Fig_MIS_performance}(a)] includes the time of 10 executions and the times of mapping and communication. The computation-time of SBM without the time of mapping is also shown there as the red dashed line. The total number of SBM executions for Fig.~\ref{Fig_MIS_performance} is 400 (10 executions per graph, 10 market graphs per universe, 4 universes).

From the computation times in Fig.~\ref{Fig_MIS_performance} (a), it is observed that SBM is the fastest for all the sizes of market graphs.
OR-tools could not output solutions for problems with 952 nodes or more by a timeout error (we define the timeout time as 10 hours).
The speed advantage of SBM becomes more pronounced as the number of nodes increases, and at 2,026 nodes, the average computation times for SBM and NetworkX are 8.54$\times10^{-2}$ and 5.32$\times10^{2}$ seconds, respectively, with SBM being 6,230 times faster.
This difference is further pronounced in practical use cases because the performance of a portfolio strategy should be analyzed for a long-term period under various conditions for optimizing strategy parameters (as we will see an example in Sec.~\ref{sec:Portfolio performance}).

The solution-accuracy (the size of the independent set found) in Fig.~\ref{Fig_MIS_performance} (b) is shown as the ratio to the largest one of the sizes of independent sets found by the three solvers. Since OR-tools is an exact-solution solver, the relative accuracy is 100\% at the node numbers 99 and 222. At the 99 and 222 nodes, the relative accuracies of the SBM are close to those of OR-tools, while NetworkX is considerably inferior to the other two solvers. More specifically, at the 99 and 222 nodes, the SBM found the exact solutions (the same solutions as ones by OR-tools) for 9 and 6 samples (out of 10 samples for each), respectively, and the sizes of the remaining solutions (1 and 4 samples for the 99 and 222 nodes, respectively) are smaller only by 1 from the sizes of the exact solutions; the SBM provide approximate solutions very close to the exact solutions. The accuracy advantage of SBM over NetworkX seen at the 99 and 222 nodes (roughly 20\%) is maintained almost constant for larger problem sizes of the 952 and 2026 nodes (OR-tools was not capable of solving those sizes of problems).

\scriptsize
\begin{table*}
\caption{Statistics of the simulated 10-year (120-month) performance data of the MIS portfolio strategy when varying $\theta$ from 0.18 to 0.36 (19 patterns) for the two cases of the asset-allocation methods (EW or IVW), including the edge density (max, min, and average) of market graphs, the size (\# of stock) of independent set found (max, min, average, and standard deviation: SD), the annualized return/risk and Sharpe ratio of the strategy.}
\label{Tbl_result1}
\fontsize{4pt}{8pt}\selectfont
\begin{tabularx}{180mm}{l|rrrrrRrrrrrrrrrrrrr} 
\toprule
\textbf{$\theta$}&0.18&0.19&0.20&0.21&0.22&0.23&0.24&0.25&0.26&0.27&0.28&0.29&0.30&0.31&0.32&0.33&0.34&0.35&0.36\\
\midrule
\textbf{Edge density}&&&&&&&&&&&&&&&&&&&\\
\quad Max (\%)&94.9&93.6&92.0&90.2&88.1&85.7&83.0&80.1&76.9&74.1&71.2&68.2&65.0&61.8&58.6&55.2&51.8&48.4&44.9\\
\quad Min (\%)&71.7&68.2&64.7&61.0&57.2&53.4&49.6&45.8&42.0&38.4&34.9&31.6&28.4&25.5&22.7&20.2&17.8&15.7&13.8\\
\quad Avr (\%)&87.5&85.4&83.2&80.6&77.9&75.0&71.9&68.6&65.2&61.7&58.1&54.5&50.8&47.1&43.5&39.9&36.5&33.1&29.9\\
\midrule
\textbf{Stock No.}&&&&&&&&&&&&&&&&&&&\\
\quad Max&132&151&173&196&225&248&278&312&350&387&420&454&501&539&593&640&680&738&784\\
\quad Min&3&27&32&37&43&52&60&71&81&96&111&129&150&166&191&218&246&276&305\\
\quad Avr&62&73&84&95&108&123&140&158&178&201&224&251&279&310&343&378&414&453&492\\
\quad SD&35&37&42&47&52&58&64&70&77&84&91&97&103&110&117&123&129&135&140\\
\midrule
\textbf{Result (EW)}&&&&&&&&&&&&&&&&&&&\\
\quad Return (\%)&18.1&18.7&16.5&16.4&17.4&16.4&16.7&16.4&15.4&16.4&15.1&15.2&15.4&15.7&14.8&14.9&15.0&15.2&15.1\\
\quad Risk (\%)&17.1&17.0&16.3&16.5&16.0&15.7&15.9&15.8&15.8&15.5&15.6&15.4&15.3&15.4&15.3&15.3&15.3&15.2&15.2\\
\quad Sharpe ratio&1.06&1.10&1.01&1.00&1.09&1.04&1.05&1.04&0.98&1.05&0.97&0.98&1.00&1.02&0.96&0.97&0.99&1.00&0.99\\
\midrule
\textbf{Result (IVW)}&&&&&&&&&&&&&&&&&&&\\
\quad Return (\%)&19.6&18.2&16.8&16.6&17.2&16.3&16.5&16.2&15.1&15.9&14.6&14.7&14.8&14.9&14.0&14.1&14.2&14.4&14.2\\
\quad Risk (\%)&15.5&15.0&14.2&14.7&14.2&14.0&14.2&14.0&14.1&13.8&13.9&13.8&13.8&13.9&13.8&13.9&13.8&13.8&13.8\\
\quad Sharpe ratio&1.26&1.21&1.18&1.13&1.21&1.16&1.16&1.16&1.07&1.15&1.05&1.06&1.07&1.08&1.01&1.02&1.03&1.04&1.03\\
\bottomrule
\end{tabularx}
\end{table*}
\normalsize

\section{Portfolio performance}\label{sec:Portfolio performance}
By using the SBM-based portfolio simulator described in Sec.~\ref{sec:Portfolio simulator}, we simulate the performance of the MIS portfolio strategy over 10 years (Apr. 1, 2013, to Mar. 31, 2023) for a large universe including 1,747 stocks corresponding to TOPIX. The strategy parameter (correlation threshold $\theta$) and the asset-allocation methods (EW or IVW) are optimized by repeating the long-term, large-scale backcast simulation. Then we compare the performance of the best MIS portfolio strategy with the Japanese major indices (assuming passive index funds) from the perspective of risk-return characteristics.

\subsection{Parameter search}\label{sec:Parameter search}

Table~\ref{Tbl_result1} shows the statistics of the simulated 10-year (120-month) performance data of the MIS portfolio strategy when varying $\theta$ from 0.18 to 0.36 (19 patterns) for the two cases of the asset-allocation methods (EW or IVW) [totally 38 settings]. As illustrated in Fig.~\ref{Fig_MIS_flowchart}, the total number of SBM executions for Table~\ref{Tbl_result1} is 22,800 (10 executions per graph, 120 market graphs per $\theta$, 19 patterns of $\theta$). For each of the 120 market graphs, the edge density and the size (the number of stocks) of the independent set found were examined and their statistics are summarized in Table~\ref{Tbl_result1}. As the correlation threshold $\theta$ increases, the number of excluded edges increases, and therefore the edges get more sparse, as a result, the size of independent set increases. These trends are consistent with the previous studies~\cite{butenko03, boginski04, boginski2005statistical}.

\begin{figure}[t]
\centering
\includegraphics[width=8.3 cm]{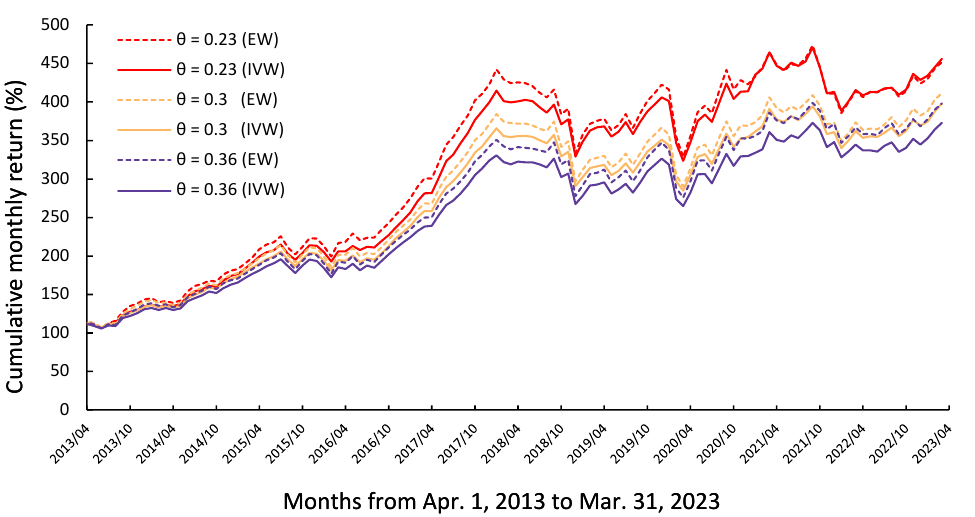}
\caption{Cumulative monthly returns of the MIS portfolio strategies with one of the two asset-allocation methods (IVW and EW) when varying the correlation threshold $\theta$. Simulation data is from Apr. 1, 2013, to Mar. 31, 2023 (10 years).}
\label{Fig_ret_performance_typ}
\end{figure}

Figure~\ref{Fig_ret_performance_typ} shows the cumulative monthly returns of the MIS portfolio strategies for the representative six settings as a function of time.
The performance of the strategy, i.e., the annualized return/risk and Sharpe ratio, are calculated from the data of the monthly returns, where the Sharpe ratio~\cite{sharpe66} is, in this work, the ratio of the mean to the standard deviation of the return (the profit and loss per period for an investment) from a strategy as in~\cite{backus93}. For both EW and IVW, the risks are reduced as $\theta$ increases. This can be interpreted as an increase in constituents of the portfolio leading to a reduction in risk, which is the diversification effect as expected. The Sharpe ratio, as well as the risk, peaks at $\theta$ = 0.18 or 0.19 and then slightly declines as $\theta$ increases, although it remains above 1 for most of the $\theta$ settings.

In this work, we select the combination of $\theta$ = 0.23 and IVW  as one that gives the best performance (Sharpe ratio = 1.16) under a constraint that the numbers of constituents in the portofolios should be more than 50 for the diversification effect. When comparing the EW and IVW methods, the annualized returns are comparable, while the annualized risks are, overall, better for the IVW method. The results demonstrate the validity of the IVW concept of predicting future risks based on past volatility.

\subsection{Performance evaluation}\label{sec:Performance evaluation}

\begin{figure}[h]
\centering
\includegraphics[width=8.3 cm]{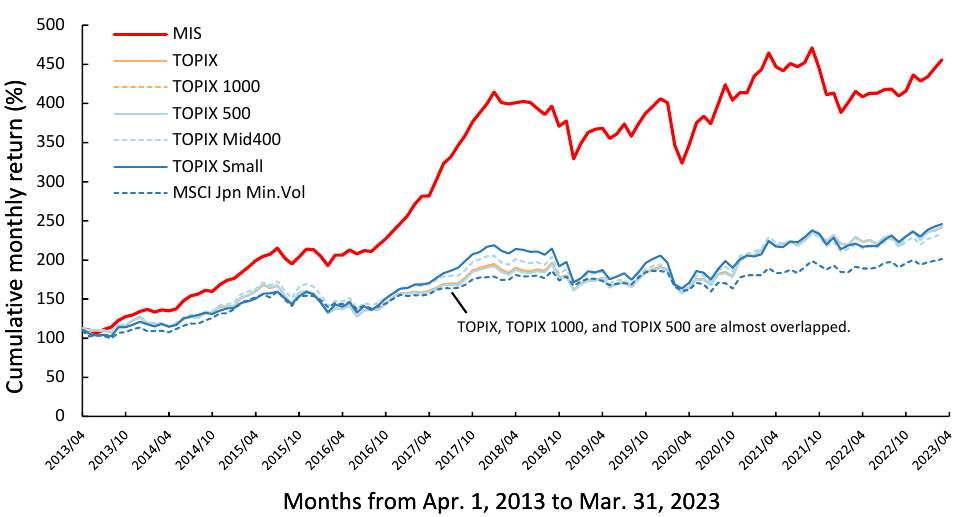}
\caption{Cumulative monthly returns of  the best MIS portfolio strategy ($\theta$ = 0.23 and IVW) and major indices in the TSE (TOPIX, TOPIX~1000, TOPIX~500, TOPIX Mid400, TOPIX Small, and MSCI Japan Minimum Volatility Index). The evaluation is for Apr.1, 2013, to Mar. 31, 2023 (10 years).}
\label{Fig_ret_performance}
\end{figure}

\scriptsize
\begin{table*}
\caption{Annualized return, risk, and Sharpe ratio of the best MIS portfolio strategy ($\theta$ = 0.23 and IVW) and major indices in the TSE (TOPIX, TOPIX~1000, TOPIX~500, TOPIX Mid400, TOPIX Small, and MSCI Japan Minimum Volatility Index). The evaluation is for Apr.1, 2013, to Mar. 31, 2023 (10 years).}
\label{Tbl_result2}
\centering
\fontsize{8pt}{12pt}\selectfont
\begin{tabular}{l|cccccc|c}
\toprule
&TOPIX&MSCI Jpn Min.Vol&TOPIX~500&TOPIX~1000&TOPIX Mid400&TOPIX Small&MIS Portfolio\\
\midrule
Return (\%)&10.0&7.7&10.0&10.0&9.7&10.2&\textbf{16.3}\\
Risk (\%)&15.2&\textbf{12.1}&15.3&15.3&15.2&15.2&14.0\\
Sharpe ratio&0.66&0.64&0.65&0.66&0.64&0.67&\textbf{1.16}\\
\bottomrule
\end{tabular}
\end{table*}
\normalsize

\begin{figure}[t]
\centering
\includegraphics[width=8.3 cm]{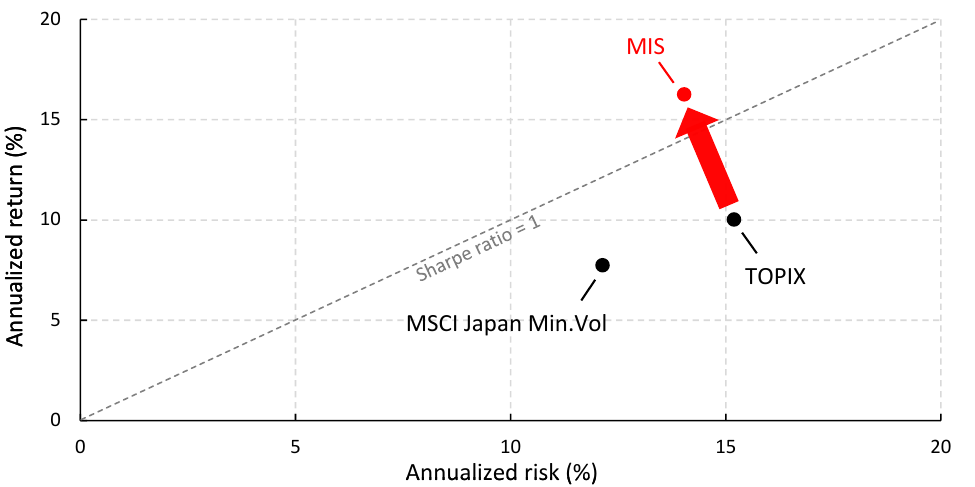}
\caption{Mapping of the best MIS portfolio strategy and major TSE indices (the data in Table~\ref{Tbl_result2}) on risk-return graph. Since the data for TOPIX~1000, TOPIX~500, TOPIX Mid400, and TOPIX Small are almost overlapped with that of TOPIX, the data for TOPIX is only illustrated.}
\label{Fig_RR_performance}
\end{figure}

Figure~\ref{Fig_ret_performance} shows the cumulative monthly returns of the best MIS portfolio strategy ($\theta$ = 0.23 and IVW) and major indices in the TSE [TOPIX, TOPIX~1000, TOPIX~500, TOPIX Mid400, TOPIX Small, and MSCI Japan Minimum Volatility Index (see APPENDIX \ref{sec:Market indices in the TSE} for the details of the indices)]. Their risk-return characteristics (annualized return/risk and Sharpe ratio) are summarized in Table~\ref{Tbl_result2} and Fig.~\ref{Fig_RR_performance}. The comparison in Table~\ref{Tbl_result2} and Fig.~\ref{Fig_RR_performance} is conservative for the MIS strategy because the trading cost is included only in the MIS strategy (see Sec.~\ref{sec:Portfolio rebalance}) [note that the dividends are considered for the MIS strategy and the major indecies].

The results clearly show that the MIS strategy outperforms all the indices in terms of annualized return and Sharpe ratio. The MIS strategy is superior to the TOPIX series indices because of the relatively high return and relatively low risk. It is remarkable that the MIS strategy (a diversified portofolio strategy) outperforms the risk-oriented index, MSCI Japan Minimum Volatility Index, in terms of the Sharpe ratio due to the relatively high return.

\begin{figure}[h]
\centering
\includegraphics[width=8.3 cm]{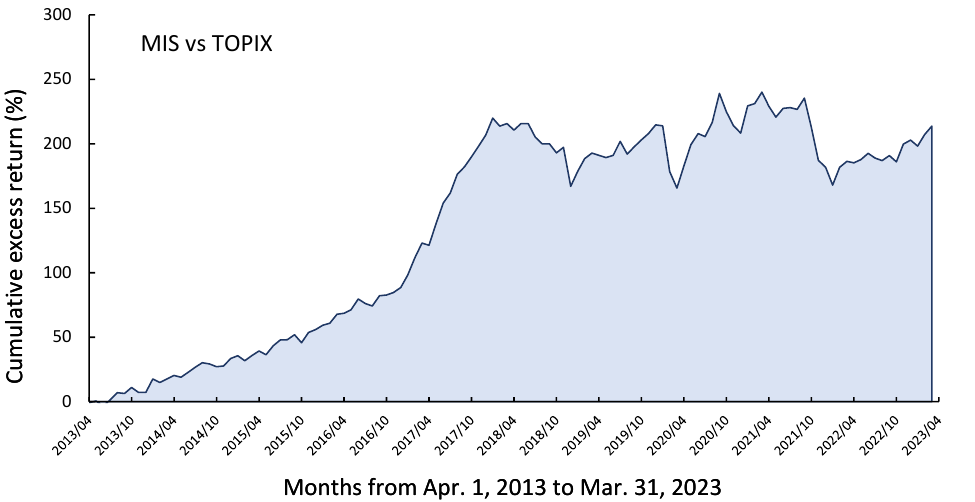}
\caption{The cumulative excess return of MIS portfolio versus TOPIX. The evaluation is for Apr.1, 2013, to Mar. 31, 2023 (10 years).}
\label{Fig_cum_excess_ret}
\end{figure}

We analyzed the factors contributing to the relatively high return of the MIS strategy. Figure~\ref{Fig_cum_excess_ret} shows the cumulative excess return for the MIS strategy versus TOPIX. The annualized return of the MIS strategy for the period from Sep. 2016 and Dec. 2017 is 49.0\%, remarkably larger than that of TOPIX (22.7\%).

To examine which and what kind of stocks contributed to the return of the MIS strategy, we introduce an index, $DIFR$, for each stock. $DIFR_{i}$ is defined as the difference in the return of stock $i$ between the MIS portfolio strategy and TOPIX for the period and expressed by
\begin{align}\label{Eq_RMIS_RTPX}
DIFR_{i}&=R^\mathrm{MIS}_{i} - R^\mathrm{TPX}_{i},\\[10pt]
R^\mathrm{MIS}_{i}&=\sum\limits_{t}^{T} {R_{i}(t)\cdot w^\mathrm{MIS}_{i}(t)},\\[10pt]
R^\mathrm{TPX}_{i}&=\sum\limits_{t}^{T} {R_{i}(t)\cdot w^\mathrm{TPX}_{i}(t)}.
\end{align}
Here, $R_{i}(t)$ is the monthly return of stock $i$ at month $t$ ($T$ is 16 months in the period). $w^\mathrm{MIS}_{i}(t)$ and $w^\mathrm{TPX}_{i}(t)$ are the weights of stock $i$ in the MIS portfolio and TOPIX at month $t$, respectively. $R^\mathrm{MIS}_{i}$ and $R^\mathrm{TPX}_{i}$ are the returns of stock $i$ in the MIS strategy and TOPIX (a TOPIX-associated passive fund), respectively.

The size of the universe for the MIS strategy is 1,747 (see Sec.~\ref{sec:Stock universe}), while the constituents of TOPIX is 2,156 as of Jul. 2023 (see APPENDIX~\ref{sec:Market indices in the TSE}). With assuming the constituents of TOPIX are the same as the universe of the MIS strategy, we calculated the performance and weights [$w^\mathrm{TPX}_{i}(t)$] for TOPIX according to the weighting method of TOPIX (a capitalization-based weighting, see APPENDIX~\ref{sec:Market indices in the TSE}). Hence, $\sum_{t=1}^{1747} w^\mathrm{MIS}_{i}(t) = 1$ and $\sum_{t=1}^{1747} w^\mathrm{TPX}_{i}(t) = 1$. We have confirmed that the performance of TOPIX for the 1,747 constituents is almost the same as that of the original TOPIX since the dominant constituents do not differ.

Table~\ref{Tbl_result3} shows the result of a deferential factor analysis of the performances between the MIS portfolio strategy and TOPIX for the period. The top five and bottom five stocks when sorted by $DIFR$ are listed there, along with the market capitalization, the degree of the node in the market graph, $w^\mathrm{MIS}$, and $w^\mathrm{TPX}$ (averaged over the period). Here, the degree of a node, in graph theory, represents the number of edges that are connected to the node.

\scriptsize
\begin{table*}
\caption{Deferential factor analysis of the performances between the MIS portfolio strategy and TOPIX.  The difference in the the return of a stock between the MIS portfolio strategy and TOPIX for a period from Sep. 2016 to Dec. 2017 is defined as $DIFR$ (see the main text). Listed are company name, stock code, market capitalization (in Japanese yen, JYN), the degree of the node in the market graph, weight in the MIS portfolio ($w^\mathrm{MIS}$), weight in TOPIX ($w^\mathrm{TPX}$) and DIFR for the top five and bottom five stocks when sorted by $DIFR$. The market capitalization, the degree of the node, $w^\mathrm{MIS}$, and $w^\mathrm{TPX}$ are averaged over the period.}
\label{Tbl_result3}
\fontsize{5pt}{8pt}\selectfont
\begin{tabular}{r|c|c|rrccc}
\toprule
Rank&
\begin{tabular}{c}Company\end{tabular}&
\begin{tabular}{c}Stock\\code\end{tabular}&
\begin{tabular}{c}Market capitalization\\(Millions of JYN)\end{tabular}&
\begin{tabular}{c}Degree of\\the node\end{tabular}&
\begin{tabular}{c}$w^\mathrm{MIS}$ (\%)\end{tabular}&
\begin{tabular}{c}$w^\mathrm{TPX}$ (\%)\end{tabular}&
\begin{tabular}{c}$DIFR$ (\%)\end{tabular}\\
\midrule
1&IK Holdings Co Ltd&2722&8,523\hspace{15pt}~&340\hspace{10pt}~&0.511&0.002&+ 2.05\\
2&Imuraya Group Co Ltd&2209&15,774\hspace{15pt}~&0\hspace{10pt}~&1.850&0.005&+ 1.91\\
3&Kawanishi Warehouse Co Ltd&9322&5,681\hspace{15pt}~&1\hspace{10pt}~&1.159&0.001&+ 1.71\\
4&Jutec Holdings Corp&3157&5,391\hspace{15pt}~&12\hspace{10pt}~&1.081&0.002&+ 1.47\\
5&Seed Co Ltd&7743&5,701\hspace{15pt}~&4\hspace{10pt}~&0.909&0.001&+ 1.46\\
\midrule
1743&Sony Group Corp&6758&4,472,465\hspace{15pt}~&1591\hspace{10pt}~&0.000&1.299&- 0.57\\
1744&Nintendo Co Ltd&7974&3,235,738\hspace{15pt}~&1325\hspace{10pt}~&0.000&0.921&- 0.57\\
1745&Keyence Corp&6861&3,673,789\hspace{15pt}~&1636\hspace{10pt}~&0.000&1.042&- 0.59\\
1746&Toyota Motor Corp&7203&12,625,742\hspace{15pt}~&1662\hspace{10pt}~&0.000&3.757&- 0.74\\
1747&Mitsubishi UFJ Financial Group Inc&8306&8,357,252\hspace{15pt}~&1591\hspace{10pt}~&0.000&2.453&- 1.04\\
\bottomrule
\end{tabular}
\end{table*}
\normalsize

The top five stocks are all small-capitalization stocks and the bottom five are all large-capitalization stocks. We have confirmed that the top five small-capitalization stocks are involved in the MIS strategy but the bottom five large-capitalization stocks are not in the MIS strategy (also see $w^\mathrm{MIS}$ in Table~\ref{Tbl_result3}). As partially shown in Table~\ref{Tbl_result3} (market capitalization and the degree of the node), it has been observed that the degree of the node tends to increase with the capitalization of the stock increases. It would be because the large-capitalization stocks have large influences on the market and are correlated with one another (associated strongly with the whole market). The stocks with the smaller degrees of the node are more likely to be selected for the MIS portfolio. Hence, the MIS strategy is more likely to be composed of small-capitalization  and low-correlation stocks. In contrast, TOPIX is composed of large-capitalization stocks due to the definition of the weighting method [TOPIX~Small is composed of small-capitalization stocks (not considering correlations)]. It is found that, in the period, the selection of small-capitalization and low-correlation stocks results in the superior return of the MIS strategy seen in Figs.~\ref{Fig_ret_performance} and \ref{Fig_RR_performance}. One of the interesting future works would be to investigate whether this phenomenon (the MIS strategy may lead to not only a low risk but also a high return) also occurs in other stock markets (such as the New York Stock Exchange and the London Stock Exchange, etc.) and other kinds of markets for other financial products.

\section{Conclusion}\label{sec:Conclusion}
We have developed a look-aside type combinatorial optimization accelerator (an Ising machine) based on a quantum-inspired parallelizable algorithm called simulated bifurcation (SB), which enables the long-term backcast simulation of a diversified portfolio strategy based on the selection of the maximum independent set in a large-scale market graph representing the correlations between stocks (the MIS problem is known to be NP-hard). We optimized the strategy parameters of the MIS portfolio strategy through iterating the long-term large-scale simulations and found that the best MIS strategy outperforms the major market indices such as TOPIX, TOPIX Small, and MSCI Japan Minimum Volatility Index.

The look-aside type accelerator has a massively-parallel custom circuit (core circuit) for the ballistic SB (a variant of the SB algorithms) featuring the time-evolution pipeline path specific to the ballistic SB (corresponding to a perfectly inelastic wall in physics) and has been implemented as a PCIe-attachable FPGA card. We also have developed the hardware abstraction layer (HAL) and the APIs (Application Programming Interface), making the SB-based FPGA accelerator accessible at a higher abstraction level to the financial simulation program running on the CPU. 

We systematically evaluated the performance of the SB-based Ising machine (fully-connection 2,048-spin size, 32-bit floating point precision) as a MIS solver using various sizes of practical market graphs (generated from the historical market data). The SB-based MIS solver provides a good approximate solution very close to the exact solution in a much shorter time than an exact-solution solver (OR-tools) and also finds the independent sets with a remarkably larger size (roughly 20\%) than a conventional heuristic solver (NetworkX). At a large-scale market graph of 2,026 nodes (too large for the exact solver to handle), the SB-based solver is 6,230 times faster than the conventional heuristic solver.

By using the SB-based solver, we investigated the performance of the MIS portfolio strategy for a large-scale universe including 1,747 stocks (corresponding to the constituents of a major market index, TOPIX) by repeating the 10-year simulation (2013 to 2023) with varying the correlation threshold parameters (19 patterns) and the asset-allocation methods (inverse volatility weight or equal weight) [totally involving the 22,800 executions of the SB solver]. By comparing with the major market indices, it has been found that the best MIS portfolio strategy based on the large-scale universe has not only a relatively low risk (the diversification effect as expected) but also a relatively high return, outperforming the major indices. We analyzed the factors contributing to the relatively high return of the MIS strategy and concluded that the selection of small-capitalization and low-correlation stocks results in the superior return.

A potential direction of further research is to examine the performance and risk-return characteristics of the MIS portfolio strategy for a large-scale universe in other stock markets or other kinds of markets for other financial products.

\section*{Appendices}
\renewcommand{\thesubsection}{\Alph{subsection}}
\subsection{Market indices in the TSE}\label{sec:Market indices in the TSE}
Table~\ref{Tbl_result4} summarizes the constituents and weighting (composition) of the major market indices in the TSE (TOPIX, TOPIX~500, TOPIX~1000, TOPIX~Mid400, TOPX~Small, MSCI Japan Minimum Volatility Index, and Nikkei~225). The numbers of component stocks are ones as of Jul. 2023.

\scriptsize
\begin{table*}
\caption{The constituents and weighting methods of the major market indices in the TSE. The numbers of component stocks are ones as of Jul. 2023. }
\label{Tbl_result4}
\fontsize{7pt}{9pt}\selectfont
\begin{tabular}{c|c|p{6cm}|p{6cm}}
\toprule
Index&Number of Stocks&Component Stocks&Weighting Method\\
\midrule
TOPIX&2,156&All companies in the prime market division of the TSE.&Weighted by free-float adjusted market capitalization which is formulated by ``\textit{total number of shares - number of shares not available for trading by public * stock price}''.\\
\midrule
TOPIX~500&498&Top 500 stocks (large and mid-cap stocks) in TOPIX by market capitalization and liquidity.&Weighted by free-float adjusted market capitalization.\\
\midrule
TOPIX~1000&993&Top 1,000 stocks in TOPIX by market capitalization and liquidity.&Weighted by free-float adjusted market capitalization.\\
\midrule
TOPIX~Mid400&398&400 stocks in the TOPIX~500, excluding the top 100 (large-cap stocks). The component stocks of the TOPIX~Mid400 are called mid-cap stocks.&Weighted by free-float adjusted market capitalization.\\
\midrule
TOPIX~Small&1,658&Stocks from TOPIX excluding TOPIX~500. The component stocks of the TOPIX~Small are called small-cap stocks.&Weighted by free-float adjusted market capitalization.\\
\midrule
MSCI Jpn Min.Vol&139&Selected from large and mid-cap stocks in the TSE according to portfolio strategy.&Weighted by using the Barra Global Equity Model\cite{menchero2010global}, for the lowest absolute risk.\\
\midrule
Nikkei~225&225&225 stocks with high liquidity in the prime market division of the TSE.&Weighted by stock price which is adjusted by the price adjustment factor.\\
\bottomrule
\end{tabular}
\end{table*}
\normalsize

\subsection{Implementation details}\label{sec:Implementation details}
The SBM-based MIS solver consists of mapping onto an Ising model by software processing (CPU) and solving the Ising problem corresponding to the MIS problem by hardware processing (FPGA). The FPGA (Intel Stratix 10 SX 2800 FPGA) on the board (Intel FPGA PAC D5005 accelerator card) has 933,120 adaptive logic modules (ALMs) including 2,753,000 adaptive look-up-tables (ALUTs, 5-input LUT equivalent) and 3,732,480 flip-flop registers, 11,721 20Kbit-size RAM blocks (BRAMs), and 5,760 digital signal processor blocks (DSPs). The SBM components in the FPGA described in Sec.~\ref{sec:SBM core circuit} were coded in a high-level synthesis (HLS) language (Intel FPGA SDK for OpenCL, ver.~19.2). The PCIe configuration of the FPGA card is PCIe Gen3x16 with a peak bandwidth of 15.75 GB/s. The software processing is executed with a single CPU [Intel Core i3-10100 (3.60 GHz, 4 cores)] and 8 GB DDR-DRAM.

The software MIS solvers of NetworkX (ver. 2.5.1) and OR-tools (ver. 9.4.1874) implemented as the Python libraries are executed on a server with dual CPUs [Intel Xeon Silver 4215R (3.20 GHz, 8 cores)] and 256 GB DDR-DRAM.

\subsection*{Acknowledgment}
The authors thank Masaya Yamasaki for the careful reading and helpful comments.

\subsection*{Conflicts of Interest}
R.H, Y.H., are K.T. are included in inventors on two U.S. patent applications related to this work filed by the Toshiba Corporation (no. 17/249353, filed 20 February 2020; no. 17/249293, filed 25 February 2021). The authors declare that they have no other competing interests.

\clearpage

\begin{figure}[h]
\vspace{0.5cm}
\noindent\includegraphics[width=1in,height=1.25in,clip,keepaspectratio]{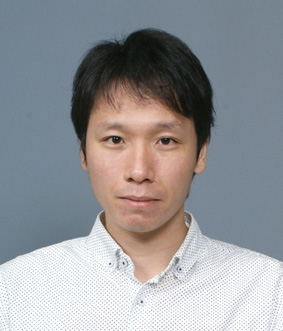}\\
{\small Ryo Hidaka received the B.E. and M.E. degrees in Systems Design and Informatics from Kyushu Institute of Technology, Japan, in 2006 and 2008, respectively. He joined Toshiba Corporation in 2008. He was engaged in the development of main processors (2D-to-3D conversion and local dimming) for digital televisions, an image recognition processor called Visconti\textsuperscript{TM}, and host controllers for flash-memory cards. His current research interests include domain-specific computing, high-level synthesis design methodology, and proof-of-concept study with FPGA devices.}
\vspace{-0.5cm} 
\end{figure}

\begin{figure}[h]
\vspace{0.5cm}
\noindent\includegraphics[width=1in,height=1.25in,clip,keepaspectratio]{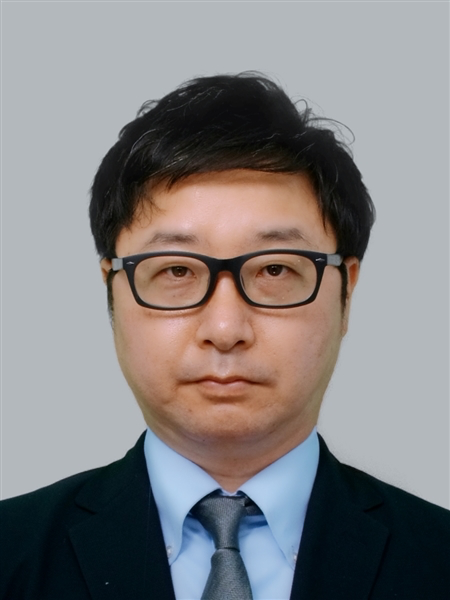}\\
{\small Jun Nakayama received the degree of Bachelor of Arts in Economic and Social studies from the University of Manchester, the U.K., in 2008. He received the degree of Master of Business Administration (MBA) in Finance from Hitotsubashi University, Japan, in 2017. He was a portfolio manager in Nomura Asset Management Co., Ltd. from 2008 to 2020 and was engaged in the development and management of quant-based funds. He joined Toshiba Corporation in 2020. He is also a Ph.D. candidate in the Financial Strategy Program, Hitotsubashi University Business School. His research interests include quantitative investment strategies, quantum-inspired computing technology, and trading strategies with advanced technologies.}
\vspace{-0.5cm} 
\end{figure}

\newpage

\begin{figure}[t]
\vspace{0.5cm}
\noindent\includegraphics[width=1in,height=1.25in,clip,keepaspectratio]{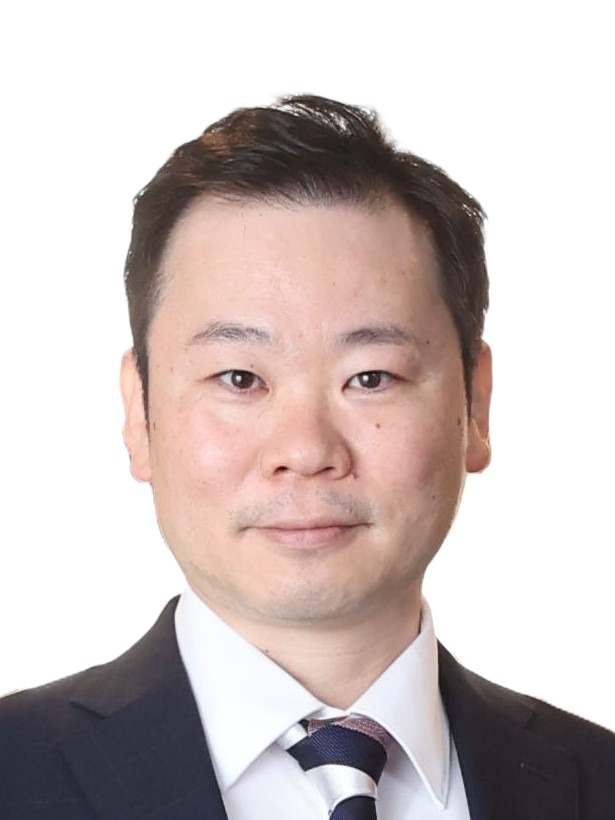}\\
{\small Yohei Hamakawa received the B.E and M.E. degrees in Advanced Electronics and Optical Science from Osaka University, Japan, in 2000 and 2002, respectively. He joined Toshiba Corporation in 2002. He was engaged in the development of image processing engines for digital televisions and TV products, and distributed computing algorithm in deep learning. His research interests include domain-specific computing, quantum computation, optimization in quantum circuit design, and their applications.}
\vspace{-0.5cm} 
\end{figure}

\begin{figure}[h]
\vspace{0.5cm}
\noindent\includegraphics[width=1in,height=1.25in,clip,keepaspectratio]{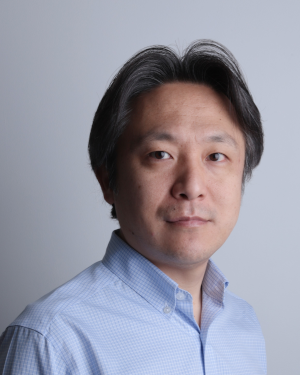}\\
{\small Kosuke Tatsumura received the B.E., M.E., and Ph.D. degrees in Electronics, Information and Communications Engineering from Waseda University, Japan, in 2000, 2001, and 2004, respectively. After working as a postdoctoral fellow at Waseda University, he joined Toshiba Corporation in 2006. He is a chief research scientist, leading a research team and several projects toward realizing innovative industrial systems based on cutting-edge computing technology. He was a member of the Emerging Research Devices (ERD) committee in the International Technology Roadmap for Semiconductors (ITRS) from 2013 to 2015. He has been a lecturer at Waseda University since 2013. He was a visiting researcher at the University of Toronto from 2015 to 2016. He received the Best Paper Award at IEEE Int. Conf. on Field-Programmable Technology (FTP) in 2016. His research interests include domain-specific computing, quantum/quantum-inspired computing, and their applications.}
\vspace{-0.5cm} 
\end{figure}

\end{document}